\documentclass[seceq]{ptptex}
\usepackage{amsmath,amssymb,amsfonts}
\usepackage{graphicx}
\usepackage{color}

\newcommand{\be}{\begin{equation}}
\newcommand{\ee}{\end{equation}}
\newcommand{\ba}{\begin{eqnarray}}
\newcommand{\ea}{\end{eqnarray}}
\newcommand \nn {\nonumber}
\DeclareMathOperator{\tr}{\mathrm{Tr}\,}
\newcommand{\ket}[1]{\left.|{#1}\right\rangle}
\newcommand{\bra}[1]{\left\langle{#1}|\right.}
\newcommand{\Zbeta}{{\cal Z}_\beta}

\title{
Towards a Theory of Entropy Production\\
in the Little and Big Bang
\footnote{arXiv:0809.4831v1}}

\author{
Teiji \textsc{Kunihiro}$^1$, 
Berndt \textsc{M\"uller}$^{2,3}$, 
Akira \textsc{Ohnishi}$^2$
and 
Andreas \textsc{Sch\"afer}$^{2,4}$
}
\inst{
$^1$Department of Physics, Kyoto University, Sakyo-ku, Kyoto 606-8502, Japan\\
$^2$Yukawa Institute of Theoretical Physics, Kyoto University, Kyoto 606-8502, Japan\\
$^3$Department of Physics, Duke University, Durham, NC 27708, USA\\
$^4$Institut f\"ur Theoretische Physik, Universit\"at Regensburg, D-93040 Regensburg, Germany
}

\abst{
We propose a broadly applicable formalism for the description of coarse grained entropy production in quantum mechanical processes. Our formalism is based on the Husimi transform of the quantum state, which encodes the notion that information about any quantum state is limited by the experimental resolution. We show in two analytically tractable cases (the decay of an unstable vacuum state and reheating after cosmic inflation) that the growth rate of the Wehrl entropy associated with the Husimi function approaches the classical Kolmogorov-Sina\"i 
entropy. We also discuss various possible applications of our formalism, 
including the production of entropy in the early stages of a relativistic 
heavy ion collision.
}

\begin{document}

\maketitle

\section{Introduction}
\label{sec:Intro}

The production of entropy at the quantum level, {\em i.~e.} the loss of information about the state of the system under consideration, is a long-standing problem in statistical physics. Many of its conceptual aspects are similar to those encountered in classical physics, but quantum mechanics poses the additional, but closely related, problem of decoherence of the quantum state of the system.  As in classical physics, one needs to distinguish two easily confused cases: (1) The loss of information about the state of a system due to its interactions with its environment, and (2) the loss of practically obtainable information due to the increasing complexity of its quantum state. The first case is by now well understood through the separation of slow and fast degrees of freedom \cite{Nakajima:1958,Zwanzig:1960,Mori:1965,Robertson:1966,Caldeira:1982iu,Balian:1986,Rau:1996ea}. The growth of entropy of the subsystem  (often involving only slow degrees of freedom and denoted by $S$) occurs because its state $|\Psi_\mathrm{S}\rangle$ becomes entangled with the state $|\Psi_\mathrm{E}\rangle$ of its environment, and the projection of the full density matrix $\hat{\rho}$ onto the system $S$ implies a loss of information described by the ``relevant'' entropy
\be
S_\mathrm{rel} = \tr \left[ \hat{\rho}_{\rm S} \ln \hat{\rho}_{\rm S} \right]
\qquad {\rm with} \qquad
\hat{\rho}_{\rm S} = \mathrm{Tr}_\mathrm{E}\, \hat{\rho}.
\label{eq:Srel}
\ee

The second case, entropy growth in an isolated quantum system, is less well understood. However, it is an important issue both in cosmology, where the early universe is thought to have made the transition from a vacuum state to a thermalized state at the end of cosmic inflation,  and in nuclear physics, where thermal matter is thought to be formed in collisions between two heavy nuclei moving at relativistic energies, both of them being in their quantum mechanical ground state. How can the evolution from a pure quantum state to a thermal ensemble be reconciled with the unitarity of the S matrix, in other words, the tendency of, e.g., the Schr\"odinger equation to preserve the purity of the initial quantum state, and how can entropy growth be described in the absence of information loss to the environment? In classical physics, the rate of information loss of nonlinear dynamical systems is described by their Kolmogorov-Sina\"i (KS) entropy, which is given by the sum over all positive Lyapunov exponents $\lambda_k>0$ of the system:
\be
S_{\rm KS} = \sum_k \lambda_k \theta(\lambda_k) .
\label{eq:SKS}
\ee
The conditions under which the KS-entropy describes the growth rate of the entropy of a classical system have been widely explored (see {\em e.~g.} \cite{Latora:1999}). There is no obvious generalization of the concept of the KS-entropy to quantum systems, whose evolution is governed by a linear equation like the Schr\"odinger equation. Nevertheless, pure quantum states are known to evolve under their own intrinsic dynamics into state which can be, for most intents and purposes, be described as thermal ensembles. Compound nuclear states are a well known example. The theory of quantum chaos \cite{QChaos} addresses the question under which conditions highly excited eigenstates of a quantum system can be approximately represented as members of a thermal ensemble \cite{Srednicki:1995pt}.

Here we are addressing a more practical question: What is the time scale on which the state of a quantum system evolves from a simple structure easily recognizable as a pure state to a complex structure which is recognized by a typical observer, who himself is limited by the uncertainty principle, as an incoherent ensemble? Can the associated growth of (apparent) entropy be defined independent of the specific details of the measurement process and how can it be calculated? The starting point of our investigation is the expectation that, like in classical mechanics, the growth of entropy is governed by the dynamics of unstable modes, which amplifies uncertainties in the initial conditions. 

One may be tempted to think that the Wigner function, which furnishes a phase-space description of quantum dynamics, provides a useful starting point to address our question. However, being not positive definite, the Wigner function does not permit a probability interpretation. Also, the Wigner function is normalized such that the occupied phase space volume essentially stays constant with time. This shows up in the fact that expanding modes always are accompanied by contracting modes (classically, the positive and negative Lyapunov exponents for a conservative system come in pairs), see Fig.1 for an example. The Husimi transform of the Wigner function \cite{Husimi:1940}, on the other hand, is positive semi-definite and admits a probability interpretation, but it also incorporates the limited ability of a typical observer to make measurements on the evolving system which can result in an increase of phase space volume, see Fig.2. It thus forms an appropriate basis for our investigation.

There are many discussions of entropy production in the literature which follow similar lines \cite{Zurek:1994wd,Pattanayak:1997,Monteoliva:2000} and address the same question for general or specific nonlinear dynamical systems. For example, Zurek and Paz \cite{Zurek:1994wd} use the example of the inverted (unstable) harmonic oscillator to show that the presence of a dissipative interaction with the environment leads to decoherence and the growth of the (von Neumann) entropy of a quantum system. Here we are mainly interested in applications to problems in nuclear and particle physics, where interactions with the environment are either weak or absent, and the growth of entropy apparent to an observer is due to the rapidly growing internal complexity of the state of an essentially closed quantum system. Thus, our approach to the problem of entropy growth is different and does not involve interactions with an environment. As far as we know, this approach as well as its application to problems in particle physics has not been described before in the literature.

In Sections \ref{sec:GrowthRate} and \ref{sec:RollOver} we investigate the Husimi function for two analytically tractable cases: the decay of an unstable quantum state and a standard  toy model of reheating (or rather a variant called pre-heating) after cosmic inflation. We show that the growth rate of the entropy associated with the Husimi function approaches the classical Kolmogorov-Sina\"i entropy in both cases. 
In Section \ref{sec:BigBang} we study entropy production during the reheating (pre-heating) 
phase of cosmic inflation.
In Section \ref{sec:Other} we compare the present entropy with other definitions such as the von Neumann entropy or the kinetic entropy.
Finally, in Section \ref{sec:RHIC}, we discuss various possible applications of our formalism, including the production of entropy in the early stages of a relativistic heavy ion collision.

\section{Entropy growth rate}
\label{sec:GrowthRate}

Following Zurek and Paz \cite{Zurek:1994wd} we choose the inverted harmonic oscillator as a simple example of an unstable mode in a quantum system:
\be
\hat{\cal H} = \frac{1}{2}\hat{p}^2 - \frac{1}{2} \lambda^2 \hat{x}^2 .
\label{eq:Ham}
\ee
Here and below we distinguish quantum mechanical operators, such as the momentum operator $\hat{p}$ from classical quantities, such as the momentum $p$, by the
caret symbol. We assume that the initial state of the system is given by a Gaussian wave packet of width $\sqrt{\hbar/\omega}$:
\be
\langle x | \psi_0 \rangle = \Bigl(\frac{\omega}{\pi\hbar}\Bigr)^{1/4} 
e^{-\omega x^2/2\hbar} .
\label{eq:init}
\ee
The Wigner function associated with the density matrix $\hat{\rho}$ is defined as
\be
W(p,x;t) = \int du \; e^{\frac{i}{\hbar}pu} \langle x - \frac{u}{2} | \; \hat{\rho}(t) \; | x + \frac{u}{2} \rangle .
\label{eq:Wdef}
\ee
It contains the full quantum mechanical information of the system and provides a phase-space picture in accordance with the uncertainty principle. The Wigner function is easily seen to satisfy the normalization condition
\be
\int \frac{dp\, dx}{2\pi\hbar} W(p,x;t) = \tr[\hat{\rho}] = 1 .
\label{eq:normW}
\ee
Similarly, one finds that
\be
\int \frac{dp\, dx}{2\pi\hbar} [W(p,x;t)]^2 = \tr[\hat{\rho}^2] \leq 1 .
\label{eq:W2}
\ee
There is no simple relationship between higher moments of the Wigner function and $\tr[\hat{\rho}^n]$ for $n>2$ . The Wigner function does not, in general, have a probabilistic interpretation, because it can take negative values as one easily confirms by calculating the Wigner function for an excited state of a harmonic oscillator.

The time evolution of $W(p,x;t)$ is determined by the equation of motion for the density matrix $\hat{\rho}$:
\be
i\hbar\frac{\partial}{\partial t} \hat{\rho}(t) = [\hat{\cal H}, \hat{\rho}(t)] .
\label{eq:rhot}
\ee
For the initial pure state, \eqref{eq:init} the density matrix is $\hat{\rho}(0)=|\psi_0\rangle\langle\psi_0|$.
As long as the Hamiltonian is at most quadratic in $x$ and $p$
or in the lowest order in $\hbar$,
the Wigner transform of Eq.~(\ref{eq:rhot}) is equivalent
to the Vlasov equation for $W$,
$\partial{W}/\partial t=\lbrace {\cal H}, W\rbrace_{P.B.}$,
where ${\cal H}(p,x)$ is the classical Hamiltonian.
Then the Wigner function is constant along the classical path,
\be
x=x_0\cosh\lambda t+p_0/\lambda \sinh\lambda t, 
\qquad
p=\lambda x_0\sinh\lambda t+p_0 \cosh\lambda t,
\ee
where $(x_0, p_0)$ are the phase space variables at $t=0$.
The time-dependent Wigner function is then easily obtained as,
\be
W(p,x;t)
= 2\, \exp\left[- \frac{1}{\hbar}
	\left(\frac{p_0^2}{\omega} + \omega x_0^2\right)\right]
= 2\, \exp\left[ - \frac{K(p,x;t)}{\hbar}\right]
\ ,
\label{eq:Wig} 
\ee
with
\ba
K(p,x;t) &=& 
\frac{p^2}{\lambda} (\sigma  \cosh{2\lambda t} + \delta)
+ \lambda x^2 (\sigma  \cosh{2\lambda t} - \delta) 
- 2\sigma\, p\, x \sinh{2\lambda t} .
\label{eq:K}
\ea
The parameters $\sigma$ and $\delta$ are defined as
\be
\sigma = \frac{\lambda^2+\omega^2}{2\lambda\omega} \geq 1,
\qquad
\delta = \frac{\lambda^2-\omega^2}{2\lambda\omega}.
\label{eq:sig}
\ee

The Husimi function is defined as the Gaussian smeared Wigner function \cite{Husimi:1940}:
\ba
H_\Delta(p,x;t) &\equiv & \int \frac{dp'\, dx'}{\pi\hbar}
\exp\left( - \frac{1}{\hbar\Delta}(p-p')^2 - \frac{\Delta}{\hbar}(x-x')^2 \right) W(p',x';t)
\nn \\
&=&
\frac{2}{\sqrt{A(t)}} \exp \left[ -\frac{1}{\hbar\, A(t)} 
\left( K(p,x;t) + \frac{p^2}{\Delta} + \Delta\, x^2 \right) \right] ,
\label{eq:Hus}
\ea
where 
\be
A(t) = 2(\sigma\rho \cosh{2\lambda t} + 1+\delta\delta')\ ,
\label{eq:At}
\ee
and the parameters $\rho$ and $\delta'$ are defined as
\be
\rho = \frac{\Delta^2+\lambda^2}{2\Delta\lambda} \geq 1\ ,
\qquad
\delta' = \frac{\Delta^2-\lambda^2}{2\Delta\lambda}\ .
\label{eq:rho}
\ee
In our convention the Husimi function is normalized according to
\be
\int \frac{dp\, dx}{2\pi\hbar} W(p,x;t) = \int \frac{dp\, dx}{2\pi\hbar} H_\Delta(p,x;t) = 1 .
\label{eq:norm}
\ee
In contrast to the Wigner function, the Husimi function is {\em always} non-negative, because it can be expressed as the expectation value of the density matrix in a coherent state:
\be
H_\Delta(p,x;t) = \langle z_\Delta | \hat{\rho}(t) | z_\Delta \rangle ,
\label{eq:Hdef}
\ee
where $\hat{a}_\Delta | z_\Delta \rangle = z_\Delta | z_\Delta \rangle$ with
\be
\hat{a}_\Delta = \frac{\Delta\,\hat{x} + i\hat{p}}{\sqrt{2\hbar\Delta}}
\label{eq:aDelta}
\ee

\begin{figure}[htb]
\centerline{
\includegraphics[width=0.46\linewidth]{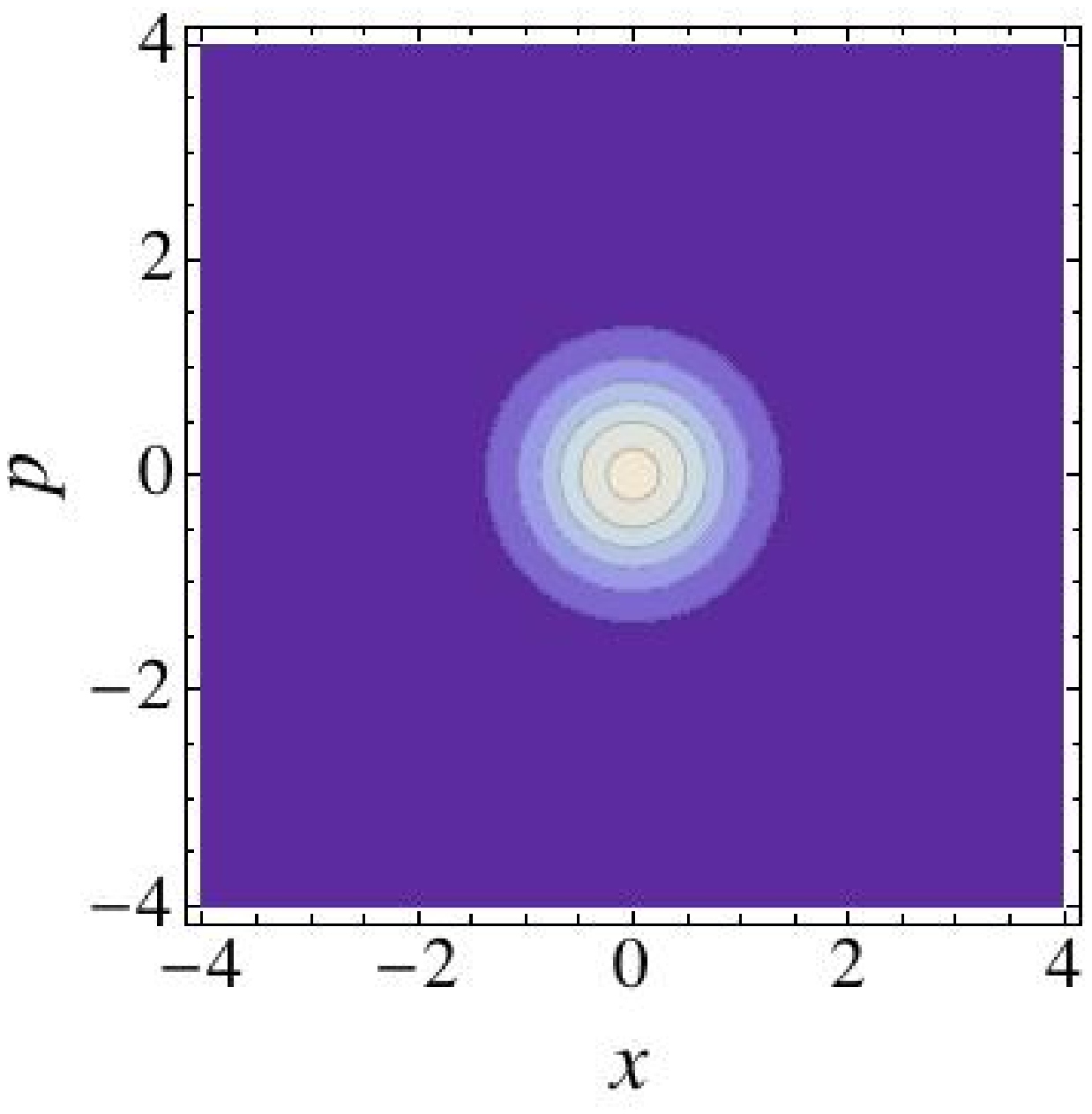}
\hspace{0.04\linewidth}
\includegraphics[width=0.46\linewidth]{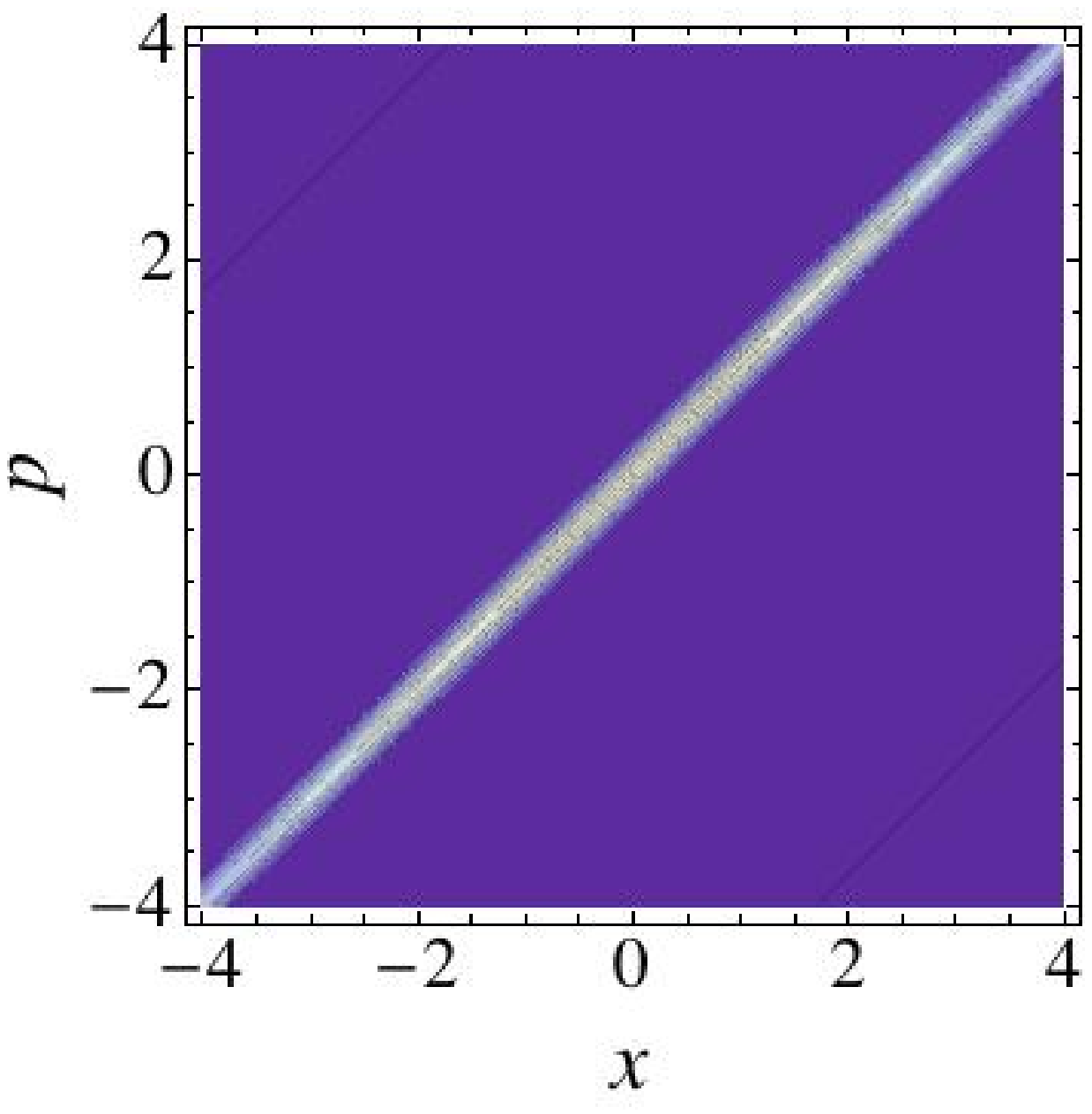}}
\caption{The Wigner function \eqref{eq:Wig} for the unstable oscillator at $t=0$ and $t=2/\lambda$ for $\sigma=1$. The horizontal axis denotes the scaled position $\omega x$; the vertical axis represents the momentum $p$.}
\label{fig1}
\end{figure}

\begin{figure}[htb]
\centerline{
\includegraphics[width=0.46\linewidth]{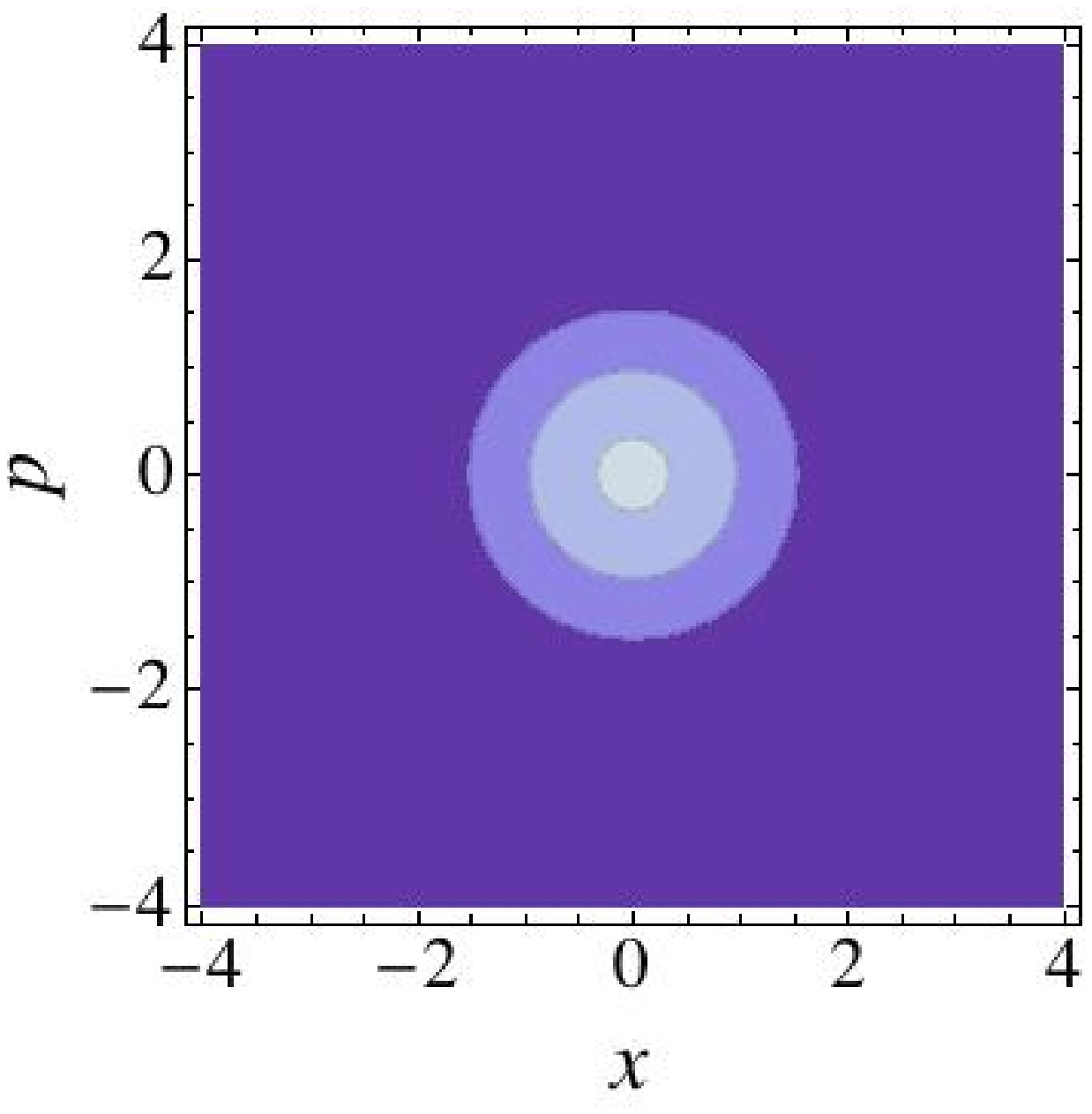}
\hspace{0.04\linewidth}
\includegraphics[width=0.46\linewidth]{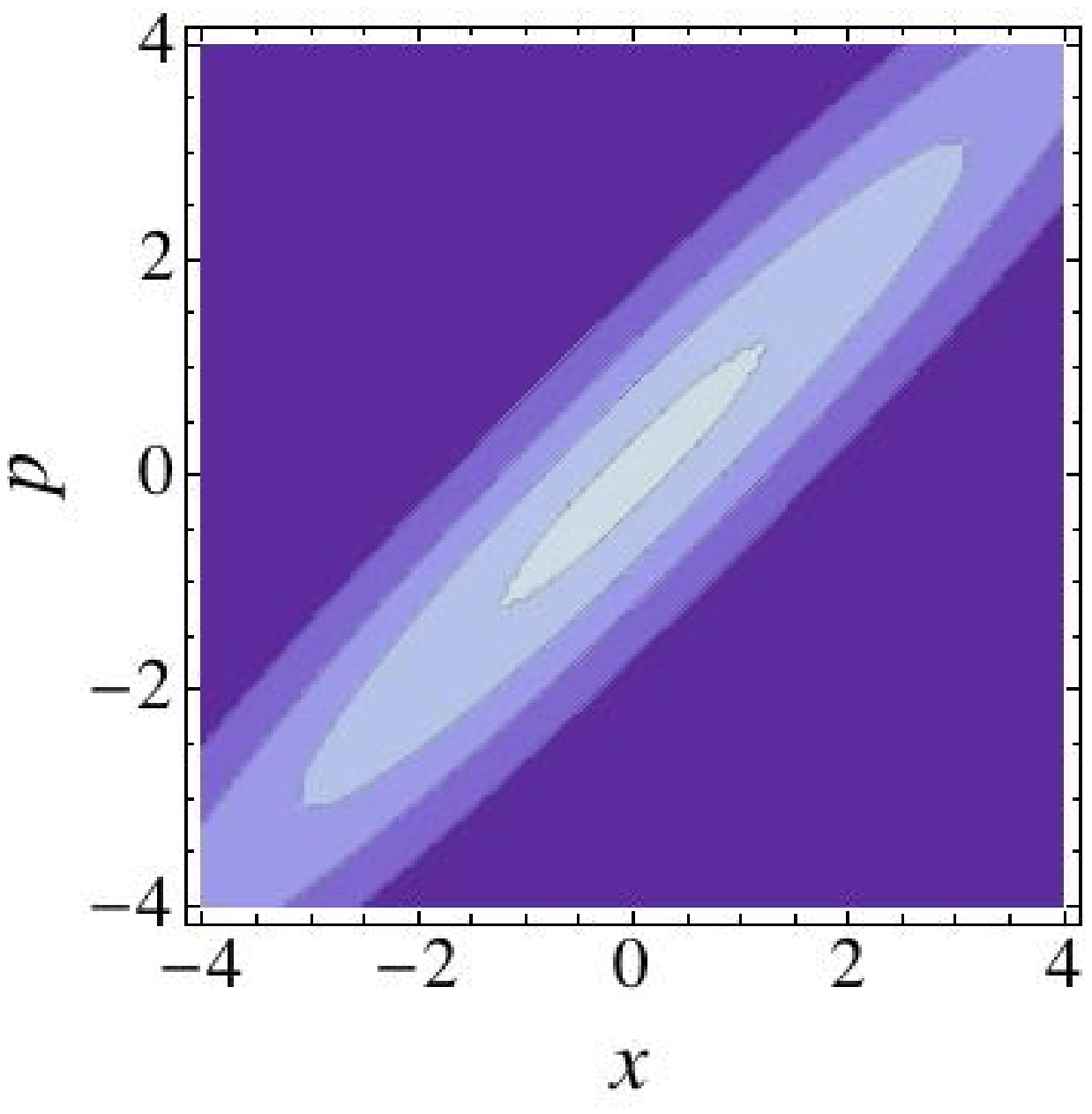}}
\caption{Husimi function \eqref{eq:Hus} for the unstable oscillator at $t=0$ and $t=2/\lambda$ for $\rho=\sigma=1$. Note that the extent of the distribution in the off-diagonal direction ($p-\omega x$) does not shrink beyond the resolution limit set by the Gaussian smearing introduced by the Husimi transform.}
\label{fig2}
\end{figure}

Because it is non-negative, the Husimi function can be used to define a 
coarse grained entropy of the quantum state, first introduced by Wehrl 
\cite{Wehrl:1978zz}:
\be 
S_{\rm H,\Delta}(t) ~=~ - \int \frac{dp\, dx}{2\pi\hbar} H_\Delta(p,x;t) \ln H_\Delta(p,x;t)
\ee
To simplify notation we introduce 
\be
L(x,p,t) ~=~ K(x,p,t) +\frac{p^2}{\Delta} +\Delta x^2
\ee
and get 
\ba
S_{\rm H,\Delta}(t) &=& 
- \int \frac{dp\, dx}{2\pi\hbar}
  \frac{2}{\sqrt{A(t)}}
  \exp\Bigl[ -\frac{L}{\hbar A(t)} \Bigr]
\Biggl( \ln \frac{2}{\sqrt{A(t)}}  - \frac{L}{\hbar A(t)} \Biggr)
\nn
\\
&=& 
\frac{1}{\hbar} 
\int \frac{dp\, dx}{2\pi} \Biggl( 
\ln \frac{\sqrt{A(t)}}{2} ~+~\hbar \frac{\partial}{\partial  \hbar} 
\Biggr)
\frac{2}{\sqrt{A(t)}} 
\exp\Bigl[ -\frac{L}{\hbar A(t)} \Bigr]
\nn
\\
&=&
\frac{1}{\hbar}
\Biggl( 
\ln \frac{\sqrt{A(t)}}{2} ~+~\hbar \frac{\partial}{\partial  \hbar} 
\Biggr)
\int \frac{dp\, dx}{2\pi} H_\Delta(p,x;t)
\nn
\\
&=&
\frac{1}{\hbar}
\Biggl( 
\ln \frac{\sqrt{A(t)}}{2} ~+~\hbar \frac{\partial}{\partial  \hbar} 
\Biggr)\hbar
\nn
\\
&=&
\ln \frac{\sqrt{A(t)}}{2} ~+~1 ~=~ \frac{1}{2}\ln \frac{A(t)}{4} + 1 
\label{eq:SH}
\ea
The Wehrl entropy is a measure of the complexity of the state of the system, as one can see as follows. The volume of support in ($2D$)-dimensional phase space of the Wigner function of a $D$-dimensional system in a pure quantum state is always equal to $h^D=(2\pi\hbar)^D$, independent of time. This property is mathematically expressed by the fact that the square of the Wigner function for any pure quantum state is equal to one:
\be
\int \frac{dp\, dx}{2\pi\hbar} W^\mathrm{(p.s.)}(p,x;t)^2 =1.
\label{eq:Wps2}
\ee
This property is a consequence of the relation 
$\tr[\hat{\rho}^2]=\tr[\hat{\rho}]=1$ valid for the density matrix of a pure quantum state. For an unstable or chaotic system, the shape of the volume in which $W(\mathbf{p},\mathbf{x};t)$ is significantly different from zero becomes more and more elongated and irregular as time progresses. If we divide phase space into a regular grid of cells of volume $h^D$, the Wigner function will take on substantially nonzero values in an increasing number of phase space cells. If we define the {\em complexity} of the state of the system as the number of cells in which the Wigner function averaged over the cell
takes values larger than a given lower bound, the complexity will grow with time as the support of the Wigner functions becomes more and more irregular. 

The Husimi function measures this growth in complexity by smearing $W(\mathbf{p},\mathbf{x};t)$ locally with a minimum uncertainty Gaussian of width $h^D$. The volume of support of the Husimi function in phase space is thus equal to the number of phase space cells ``touched'' by the Wigner function in a significant way. The growth in the volume of support of the Husimi function in phase space can be expressed in terms of its second moment
\be
M_2 \equiv \int \frac{dp\, dx}{2\pi\hbar} H_\Delta^\mathrm{(p.s.)}(p,x;t)^2 \leq 1.
\label{eq:H2}
\ee
The inverse of $M_2$ has been proposed as a convenient and easily calculable measure of the complexity of the state of a quantum system \cite{Sugita:2001ks,Sugita:2001hw}. Its great disadvantage, for our purposes, is that it is not a measure of the coarse grained entropy of a quantum system and does not approach the von Neumann entropy for a system described by a thermal ensemble. The Wehrl entropy, $S_{\rm H,\Delta}$, on the other hand, shares in the ability of the Husimi function to trace the complexity of a quantum state, and it permits an interpretation as the entropy of a quantum system. We thus focus our investigation on it.

The rate of growth of the Wehrl entropy
\ba
\frac{dS_{\rm H,\Delta}}{dt} 
&=& \int \frac{dp\, dx}{2\pi\hbar} \frac{\partial H_\Delta}{\partial t} \ln H_\Delta
+ \frac{\partial}{\partial t}\int \frac{dp\, dx}{2\pi\hbar} H_\Delta
=
\int \frac{dp\, dx}{2\pi\hbar} \frac{\partial H_\Delta}{\partial t} \ln H_\Delta
\nn \\
&=& \frac{\lambda\, \sigma\rho\, \sinh{2\lambda t}}{\sigma\rho\, \cosh{2\lambda t} + 1 + \delta\delta'}
\buildrel{t\to\infty}\over{\longrightarrow} \lambda
\label{eq:dSHdt}
\ea
asymptotically tends to the growth rate of the unstable mode, {\em i.e.} to the positive Lyapunov exponent of the classical Hamiltonian. This result has several noteworthy features. First, the entropy growth rate tends to the classical Lyapunov exponent very rapidly, on the time scale of $\lambda^{-1}$ itself. Second, the value of the growth rate for $t\gg\lambda^{-1}$ is independent of the smearing parameter $\Delta$. Although the absolute value of the Wehrl entropy \eqref{eq:SH} depends logarithmically on $\Delta$ via $\rho$ and $\delta'$, its growth rate does not.

Let us repeat and expand a bit the basic physics idea behind the analysis just presented: Any kind of measurement results in some coarse graining, which in turn can result in entropy production for a quantum system. The question is whether one can define entropy independently of the specifics of a measurement, only taking into account the principal limitations imposed on any measurement by quantum mechanics. The Husimi function provides an answer, as it incorporates the fact that any measurement (where here we only regard those which can be parametrized by a width $\Delta$) must fulfill the uncertainty principle. In general, the value of the coarse grained entropy will depend on the value of $\Delta$ and thus will be ambiguous. The central result obtained here is that the entropy growth rate at sufficiently long, but not too long, times is independent of $\Delta$. We conjecture, but have not proved, that the entropy growth rate is independent of other parameters one could introduce to specify an actual measurement and that the quantity $dS_{\rm H,\Delta}/dt$ is a well-defined, physical property of the system and not of the measurement process.
We note that the connection between the exponential rate of growth of the phase space volume occupied by the Husimi function and the classical Lyapunov exponents was pointed out before, {\em e.~g.}, by Toda and Ikeda \cite{Toda:1987}. These authors, however, did not draw the connection between the growth rate of the Wehrl entropy and the Lyapunov exponents of the classical theory.

Next we consider a system with a large (possibly infinite) number of unstable modes. Assuming no interactions among these modes, the Hamiltonian can be written as
\be
\hat{\cal H} = \sum_k \frac{1}{2} \left( \hat{p}_k^2 - \lambda_k^2 \hat{x}_k^2 \right)
\label{eq:Ham-N}
\ee
Assuming that the initial state is a product of Gaussians of the form 
of Eq.(\ref{eq:init})
the Wigner function and the Husimi function of the complete system are given by
\ba
W(\{p_k\},\{x_k\};t) &=& \prod_k W^{(k)}(p_k,x_k;t)
\\
H_\Delta(\{p_k\},\{x_k\};t) &=& \prod_k H_\Delta^{(k)}(p_k,x_k;t)
\label{eq:WH-N}
\ea
Using the normalization condition for the Husimi functions of the individual modes $H_\Delta^{(k)}$ it is now easy to show that
\be
S_{\rm H,\Delta}(t) = \sum_k S_{\rm H,\Delta}^{(k)}(t)
\label{eq:SH-N}
\ee
The growth rate of the total Wehrl entropy is then
\be
\frac{dS_{\rm H,\Delta}}{dt} = \sum_k \frac{\lambda_k\, \sinh{2\lambda_k t}}{\cosh{2\lambda_k t} + (1+\delta\delta')\sigma^{-1}\rho^{-1}}
\label{eq:dSHdt-N}
\ee
which, for long times, tends to the Kolmogorov-Sina\"i entropy of the system:
\be
\frac{dS_{\rm H,\Delta}}{dt} \buildrel{t\to\infty}\over{\longrightarrow} \sum_k \lambda_k .
\label{eq:KS}
\ee
The limit is approached in such a way that the large contributions (large $\lambda_k$) are reached rapidly while the small ones take longer. The dependence on the details of the Gaussian smearing disappears  exponentially with time.

\section{``Roll-over'' transition}
\label{sec:RollOver}

We next consider a system with infinitely many degrees of freedom, namely, relativistic quantum fields with dynamically unstable modes. Here we will see that, while the entropy grows linearly with time for unstable modes, it oscillates around a constant for stable modes. We will also find that the entropy growth for a translationally invariant instability rate is an extensive quantity, {\em i.~e.} it grows with the volume. 

For the purpose of analytical tractability we choose the scalar field in (1+1) dimensions with the tachyonic Lagrangian (we set $\hbar=1$ in this section.) 
\be
{\cal L} = \int dx \frac{1}{2} \left[ \left( \frac{\partial\Phi}{\partial t} \right)^2 - \left( \frac{\partial\Phi}{\partial x} \right)^2 + \mu^2\Phi^2 \right].
\label{eq:Lphi}
\ee
The modes with momentum $|p|<\mu$ are unstable, {\em i.~e.} their amplitudes 
grow exponentially with time; the modes with $|p|>\mu$ are stable and exhibit 
oscillatory behavior. The Wigner functional $W[\Pi(x),\Phi(x)]$ for the field 
$\Phi$ and its canonical momentum $\Pi=\partial\Phi/\partial t$ can be defined in complete analogy to the Wigner function of a quantum system with many degrees of freedom \cite{Mrowczynski:1994nf} as:
\ba
W[\Pi (x), \Phi (x); t]
&=& \int {\cal D}\varphi(x) \; e^{ -i\int dx \;\Pi (x) \varphi(x) }
\nn \\
&& 
\times \langle \Phi (x) + \frac{1}{2} \varphi(x) \vert \; \hat{\rho}(t) \; 
\vert \Phi(x) - \frac{1}{2} \varphi(x) \rangle 
\label{eq:WF1}
\ea

The Wigner functional for $t>0$ can be represented in terms of the Fourier modes $\Pi_p,\Phi_p$. We note that the Wigner functional is a constant along the classical trajectory $(\Phi_p(t),\Pi_p(t))$:
\be
W[\Pi,\Phi;t]=W\left[\{\Pi_p^0\},\{\Phi_p^0\},t=0\right]
\ee
where 
$\Phi_p^0$ and $\Pi_p^0$ are the initial Fourier components. If we assume that at $t=0$ the scalar field is in the vacuum state corresponding to mass $m$, this property allows to write the Wigner functional as:
\be
W[\Pi,\Phi;t] =   C \; 
\exp\left[ - \int \frac{dp}{2\pi} \;
\left( \frac{|\Pi_p^0|^2}{E_p} + E_p |\Phi_p^0|^2 \right)
\right]
\label{eq:WF2}
\ee
where for $0< p < \mu$: 
\ba
\Phi_p^0 & = &\Phi_p(t) \cosh{\lambda_p t} - 
\frac{\Pi_p(t)}{\lambda_p} \sinh{\lambda_p t} 
\label{eq:Phi0} 
\\
\Pi_p^0 & = &\Pi_p(t) \cosh{\lambda_p t} - 
\lambda_p \; \Phi_p(t) \sinh{\lambda_p t} 
\label{eq:Pi0}
\ea
with $\lambda_p=\sqrt{\mu^2-p^2}$, and for $p > \mu$:
\ba
\Phi_p^0 & = &\Phi_p(t) \cos{\omega_p t} - 
\frac{\Pi_p(t)}{\omega_p} \sin{\omega_p t} 
\label{eq:Phi0s} 
\\
\Pi_p^0 & = &\Pi_p(t) \cos{\omega_p t} + 
\omega_p \; \Phi_p(t) \sin{\omega_p t} 
\label{eq:Pi0s}
\\
\ea
with $E_p = \sqrt{p^2+m^2}$ and $\omega_p=\sqrt{p^2-\mu^2}$.

\begin{figure}[htb]
\centerline{
\includegraphics[width=0.68\linewidth]{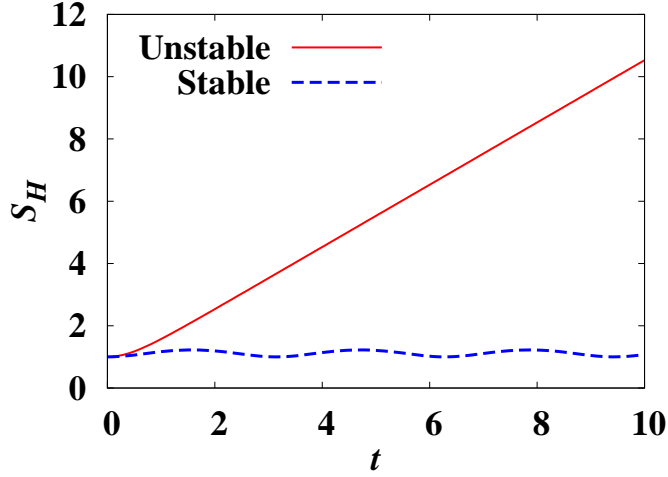}}
\caption{Contribution of an unstable mode, shown as solid line to the Wehrl entropy compared with the contribution of a stable mode, shown as dashed line. The contribution of the unstable mode grows linearly with time, while the contribution of the stable mode oscillates around a constant value. The curves are for the parameters $\lambda_p=\omega_p=1$ and $E_p=\Delta=2$. }
\label{fig4}
\end{figure}

The evaluation of the Wigner and Husimi functions proceeds in complete analogy with the case of a single unstable mode discussed in the previous section. The final result is:
\be 
H_\Delta[\Pi,\Phi;t] = \prod_{|p|<\mu} \frac{2}{\sqrt{A_p(t)}} \exp \left[ - \frac{R(\Pi_p,\Phi_p;t)}{ A_p(t)} \right]
\times \prod_{|p|>\mu} \frac{2}{\sqrt{\tilde{A}_p(t)}} \exp \left[ - \frac{\tilde{R}(\Pi_p,\Phi_p;t)}{\tilde{A}_p(t)} \right]
\label{eq:HDt}
\ee
with
\ba
R(\Pi_p,\Phi_p;t) &=&  \frac{|\Pi_p|^2}{\lambda_p} (\sigma_p  \cosh{2\lambda_p t} + \delta_p)
+ \lambda_p |\Phi_p|^2 (\sigma_p  \cosh{2\lambda_p t} - \delta_p) 
\nn \\
&&- \sigma_p (\Pi_p^*\Phi_p + \Pi_p\Phi_p^*) \sinh{2\lambda_p t} 
 + \frac{|\Pi_p|^2}{\Delta} +\Delta|\Phi_p|^2 ,
\nn \\
\tilde{R}(\Pi_p,\Phi_p;t) &=&  \frac{|\Pi_p|^2}{\omega_p} (\tilde\sigma_p + \tilde\delta_p \cos{2\omega_p t} )
+ \omega_p |\Phi_p|^2 (\tilde\sigma_p - \tilde\delta_p  \cos{2\omega_p t}) 
\nn \\
&&+ \tilde\delta_p (\Pi_p^*\Phi_p + \Pi_p\Phi_p^*) \sin{2\omega_p t} 
+ \frac{|\Pi_p|^2}{\Delta} +\Delta|\Phi_p|^2 ;
\label{eq:tIp}
\ea
\ba
A_p(t) &=& \frac{\Delta^2+\lambda_p^2}{\lambda_p\Delta}
\,{\sigma_p}\, 
\cosh{2\lambda_p t} 
+ 2 + \delta_p \frac{\Delta^2-\lambda_p^2}{\lambda_p\Delta} ,
\nn \\
\tilde{A}_p(t) &=& \frac{\Delta^2+\omega_p^2}{\omega_p\Delta}
\,\tilde{\sigma}_p\,
+ 2 +\tilde{\delta}_p \frac{\Delta^2-\omega_p^2}{\omega_p\Delta} \cos{2\omega_p t} 
\label{eq:tAp}
\ea
with 
\ba
\sigma_p = \frac{\lambda_p^2+E_p^2}{2\lambda_p E_p} \geq 1,
\qquad
\delta_p = \frac{\lambda_p^2-E_p^2}{2\lambda_p E_p} ,
\nn \\
\tilde\sigma_p = \frac{\omega_p^2+E_p^2}{2\omega_p E_p} \geq 1,
\qquad
\tilde\delta_p = \frac{\omega_p^2-E_p^2}{2\omega_p E_p}.
\label{eq:sigp}
\ea
Like in the single mode case, the Wehrl entropy is given by
\ba
S_{\rm H,\Delta}(t) &=&  \int \frac{D\Pi\, D\Phi}{2\pi}\, H_\Delta \ln H_\Delta
\nn \\
&=& V \int_{|p|<\mu} \frac{dp}{2\pi} \left[ \frac{1}{2} \ln \frac{A_p(t)}{4} + 1 \right]
+ V \int_{|p|>\mu} \frac{dp}{2\pi} \left[ \frac{1}{2} \ln \frac{\tilde{A}_p(t)}{4} + 1 \right]
\label{eq:SHp} ,
\ea
where $V$ is the quantization volume. In analogy to \eqref{eq:dSHdt}, the growth rate of the entropy is given by
\ba
\frac{dS_{\rm H,\Delta}}{dt} 
&=& V \int_{|p|<\mu} \frac{dp}{2\pi} 
\frac{\sigma_p(\Delta^2+\lambda_p^2)\sinh{2\lambda_p t}}{A_p(t)\Delta}
+ V \int_{|p|>\mu} \frac{dp}{2\pi} 
\frac{\tilde\delta_p(\omega_p^2-\Delta^2)\sin{2\omega_p t}}{\tilde{A}_p(t)\Delta}
\nn \\
&&\buildrel{t\to\infty}\over{\longrightarrow}\ V \int_{-\mu}^{\mu} \frac{dp}{2\pi}\, \lambda_p 
= \frac{V\,\mu^2}{4} .
\label{eq:dSHdtp}
\ea

It is instructive to compare the integrands in \eqref{eq:SHp} for a typical stable and unstable mode. Figure~\ref{fig4} shows such a comparison. One clearly sees that the contribution to $S_{\rm H,\Delta}$ of the unstable mode (shown as solid line) grows linearly with time, while the contribution of the stable mode (shown as dashed curve) does not grow and only oscillates slightly around its initial value. It is also noteworthy that the rate of entropy growth \eqref{eq:dSHdtp} is proportional to the volume $V$, implying that the entropy growth rate is an extensive quantity. This means that the evolution of the scalar field is characterized by a finite growth rate of the entropy density $S/V$. Generally, the entropy growth rate will be an extensive quantity, if all modes below a certain momentum scale are dynamically unstable.

\section{Entropy production in the ``Big Bang''}
\label{sec:BigBang}

A different case where the dynamics of the system is thought to lead to coarse grained entropy production is the ``(p)re-heating'' after inflation problem \cite{Bassett:2005xm}. In this case one often considers the following toy model: A scalar field $\chi$ (representing matter degrees of freedom) interacts with the inflaton field $\Phi$ via the Lagrangian
\be
{\cal L}(\hat{\chi}) 
= \frac{1}{2} \left( g^{\mu\nu}\frac{\partial \hat{\chi}}{\partial x^\mu}\, \frac{\partial \hat{\chi}}{\partial x^\nu}  
- g^2 \Phi(t)^2 \hat{\chi}^2 \right)
\label{eq:Lchi}
\ee
Because the inflaton field has the same value everywhere in space, the problem is translationally invariant and can be separated into independent modes by using the Fourier expansion (we set $\hbar=1$ in this section):
\be
\hat{\chi}(\mathbf{x},t) =  \frac{1}{R(t)} \int \frac{d^3k}{(2\pi)^{3/2}} \left( \hat{X}_k(t) e^{i\mathbf{k}\cdot\mathbf{x}} + \hat{X}^\dagger_k(t) e^{-i\mathbf{k}\cdot\mathbf{x}} \right)
\label{eq:chik}
\ee
where $R(t)$ is the scale factor describing the expansion of the universe. After inflation, the inflaton field $\Phi$ oscillates with a large amplitude and can be described, to good approximation, as a classical field:
\be
\Phi(t) \approx \Phi_0 \cos(\omega t)
\label{eq:Phit}
\ee
After rescaling the physical time variable as $\tau = \sqrt{t/t_c}$ with a characteristic cosmological time constant $t_c$, the equation for the quantum operators describing the individual Fourier modes of the matter field then takes the form of a Mathieu equation (see {\em e.g.} \cite{Greene:1997fu}):
\be 
\frac{\partial^2 \hat{X}_k}{\partial\tau^2} 
+ \left( \bar\kappa^2 + \frac{g^2}{2\lambda} \cos(2\omega\tau) \right) \hat{X}_k(\tau) = 0 ,
\label{eq:Xk}
\ee
where $\lambda$ is the self-interaction constant of the inflaton field
and $\bar\kappa^2=k^2/(\lambda\Phi_0^2)$. 
The solutions of this equation are the Mathieu sine and cosine functions $S(a,q;\omega\tau)$ and $C(a,q;\omega\tau)$ with $a=\bar\kappa^2$ and $q=-g^2/4\lambda$. We will henceforth drop the parameters $a$ and $q$. Asymptotically, the Mathieu functions can be represented as
\ba
C(\omega\tau) &\approx & e^{\mu\tau} \cos(\omega\tau + \alpha_c(\tau))
\ ,\quad
S(\omega\tau) 
\approx
e^{\mu\tau} \cos(\omega\tau + \alpha_s(\tau))
\label{eq:Mathieu}
\ea
with $\mu\geq 0$ and $\alpha_c(0)=0$, $\alpha_s(0)=-\pi/2$. For large times, the constancy of the Wronskian $(C \dot{S}-S\dot{C})/\omega=1$ ensures that the phases of both solutions approach each other with exponential precision, according to
\be
| \alpha_c(\tau) - \alpha_s(\tau) | \buildrel{\tau\to\infty}\over{\longrightarrow} e^{-2\mu\tau} .
\label{eq:dphi}
\ee

In order to explore the consequences of this contraction in phase angle, we choose the phase-amplitude representation of phase space for each mode: $(X_k,P_k)=(r_k\cos{\alpha_k},\omega r_k\sin{\alpha_k})$. Thus we have in this case as conjugate variables the squared amplitude $r_k^2/2$ and the phase $\alpha_k$. Because in our case, both $X_k(\tau)$ and $P_k(\tau)$ asymptotically oscillate with the same phase $\alpha_c(\tau) \approx \alpha_s(\tau)$, the uncertainty of the particle number (see Eq.~(238) in \cite{Bassett:2005xm})
\be
N_k \sim \frac{1}{2} \left( \frac{P_k^2}{\omega^2} + X_k^2 \right)
= \frac{r_k^2}{2} \cos^2 \alpha(\tau)
\label{eq:Nk}
\ee
is equal to $N_k$ rather than $\sqrt{N_k}$. Thus we have 
\be
(\Delta N_k)^2(\Delta \alpha_k)^2 = N_k^2(\Delta \alpha_k)^2 \ge \frac{1}{4} .
\label{eq:Nalpha}
\ee
Although it is quite commonly used, the status of \eqref{eq:Nalpha} is not unproblematic (see, {\em e.g.}, the discussion in \cite{Jackiw:1968}). 
In general, the expectation value of the commutator 
$[\hat N_k,\hat \alpha_k]$ is state-dependent. 
Our states (the solutions $\hat X_k$ of the 
Mathieu equation) are characterized completely by $N_k$ and
$\alpha_k$ and for them Eq.(\ref{eq:Nalpha}) holds. Thus we here regard 
$N_k$ and $\alpha_k$ as canonical conjugate for the states we consider. 
As a result Eq.(\ref{eq:Nalpha}) generates coarse-graining, which in turn 
can be linked to entropy production. Therefore, the Wehrl entropy defined below
is a sensible way to define entropy for our setting.

We also note that the $2\pi$-periodicity of $\alpha_k$ implies that $r_k^2/2$ is an integer. However, for large enough $\tau$, this quantization becomes irrelevant, and $r_k^2$ can be treated as a continuous variable. 

In the following we drop the mode index $k$. We assume that the wave function $\Psi(\alpha,\tau)$ is nearly independent of the phase angle initially and then contracts according to \eqref{eq:dphi}:
\be
\Psi(\alpha,\tau) \approx 
\left(\tilde\pi(\tau)\sqrt{\pi}\right)^{-1/2} \exp\left(-\frac{\alpha^2}{2\tilde\pi(\tau)^2}\right) ,
\label{eq:Psialpha}
\ee
where $\tilde\pi(\tau) = \pi \exp(-2\mu\tau)$. We shall drop the argument of $\tilde\pi$ in the following, but we assume $\tau$ to be large enough so that $\tilde\pi(\tau) \ll \pi$. It is now a straightforward exercise to calculate the Wigner function:

\ba
W(\alpha,n,\tau)
&=& \int_{-\pi}^{\pi}\, d\beta\, e^{in\beta}\,
\Psi\left(\alpha-\frac{\beta}{2},\tau\right) \Psi^*\left(\alpha+\frac{\beta}{2},\tau\right)
\nn\\
&=& 
\frac{e^{-\alpha^2/\tilde\pi^2}}{\tilde\pi\sqrt{\pi}}
 \int_{-\pi}^{\pi} d\beta e^{in\beta} \exp\left(-\frac{\beta^2}{4\tilde\pi^2}\right)
\approx
2 e^{-\alpha^2/\tilde\pi^2 - n^2\tilde\pi^2} ,
\label{eq:Walphar}
\ea
where $n \equiv r^2/2$ is an integer.  For continuous variables  
$r$ would  be $e^{\mu\tau}$. The Husimi function is similarly obtained as
\ba
H_\Delta(\alpha,n,\tau) &=& \int_{-\pi}^{\pi} \frac{d\alpha'}{\pi} \sum_{n'}
e^{-\Delta(\alpha-\alpha')^2-\frac{(n-n')^2}{\Delta}} W(\alpha',n',\tau)
\nn\\
&\approx&
\frac{2\sqrt{\tilde\pi^2\Delta}}{1+\tilde\pi^2\Delta} 
\exp\left(-\frac{\alpha^2\Delta + n^2\tilde\pi^2}{1+\tilde\pi^2\Delta} \right)
\label{eq:Halphar}
\ea

\begin{figure}[htb]
\centerline{
\includegraphics[width=0.46\linewidth]{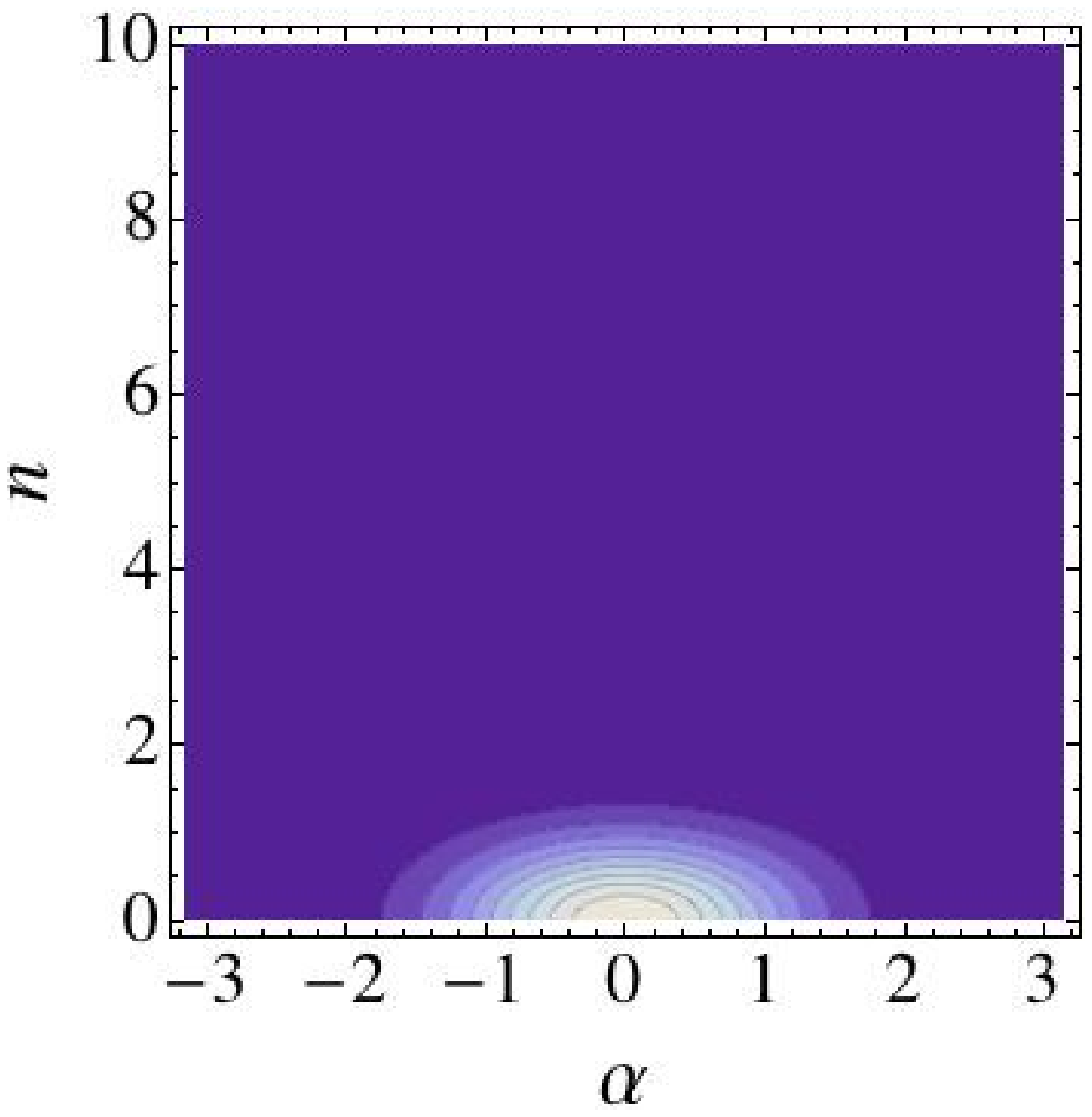}
\hspace{0.04\linewidth}
\includegraphics[width=0.46\linewidth]{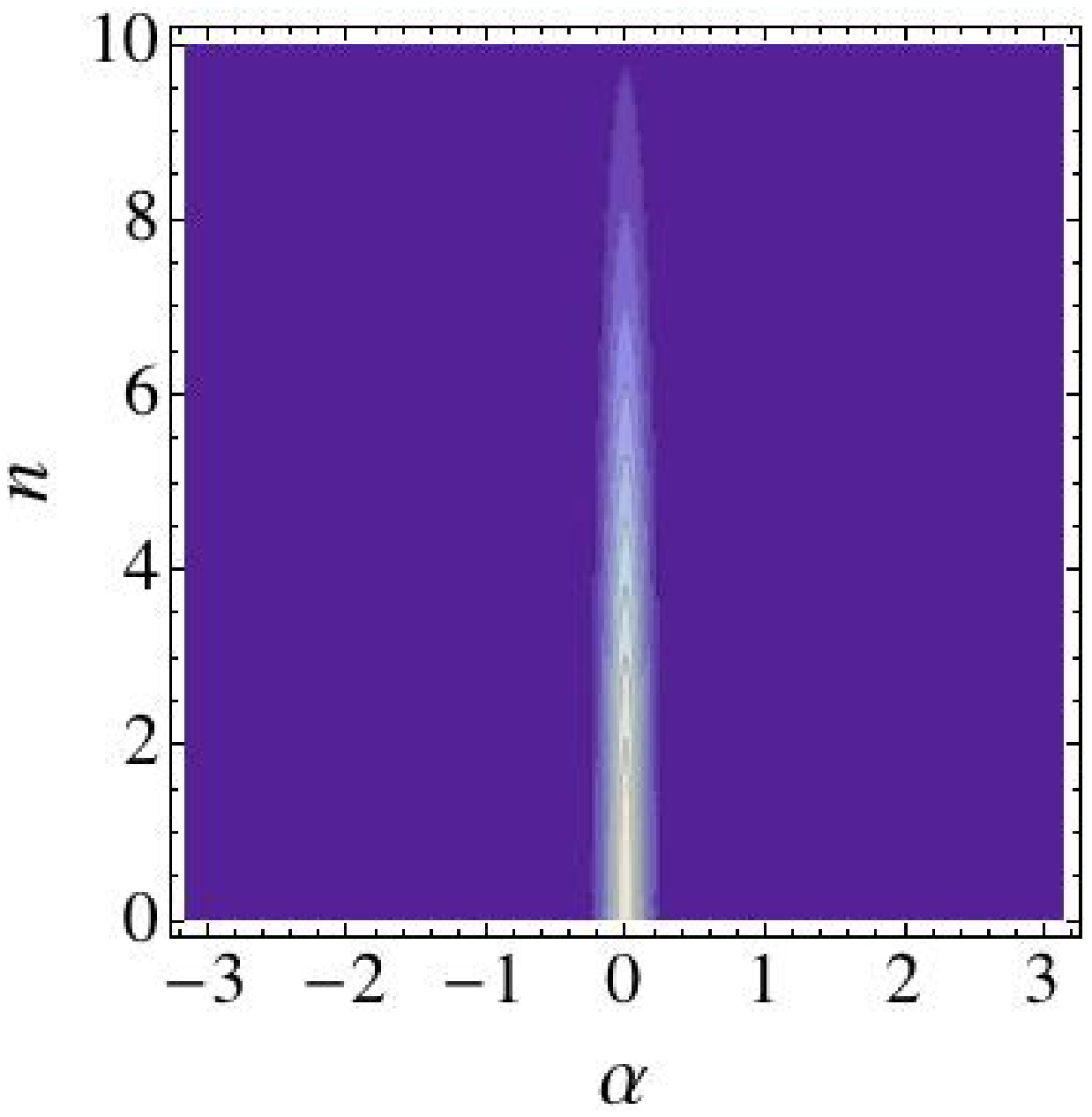}}
\caption{Wigner function \eqref{eq:Walphar} for the inflationary reheating model at $\mu\tau=0.5, 1.5$. The horizontal axis denotes the phase angle $\alpha \in [-\pi,\pi]$; the vertical axis represents the squared amplitude $r^2/2 \in [0,10]$.}
\label{fig5}
\end{figure}

\begin{figure}[htb]
\centerline{
\includegraphics[width=0.46\linewidth]{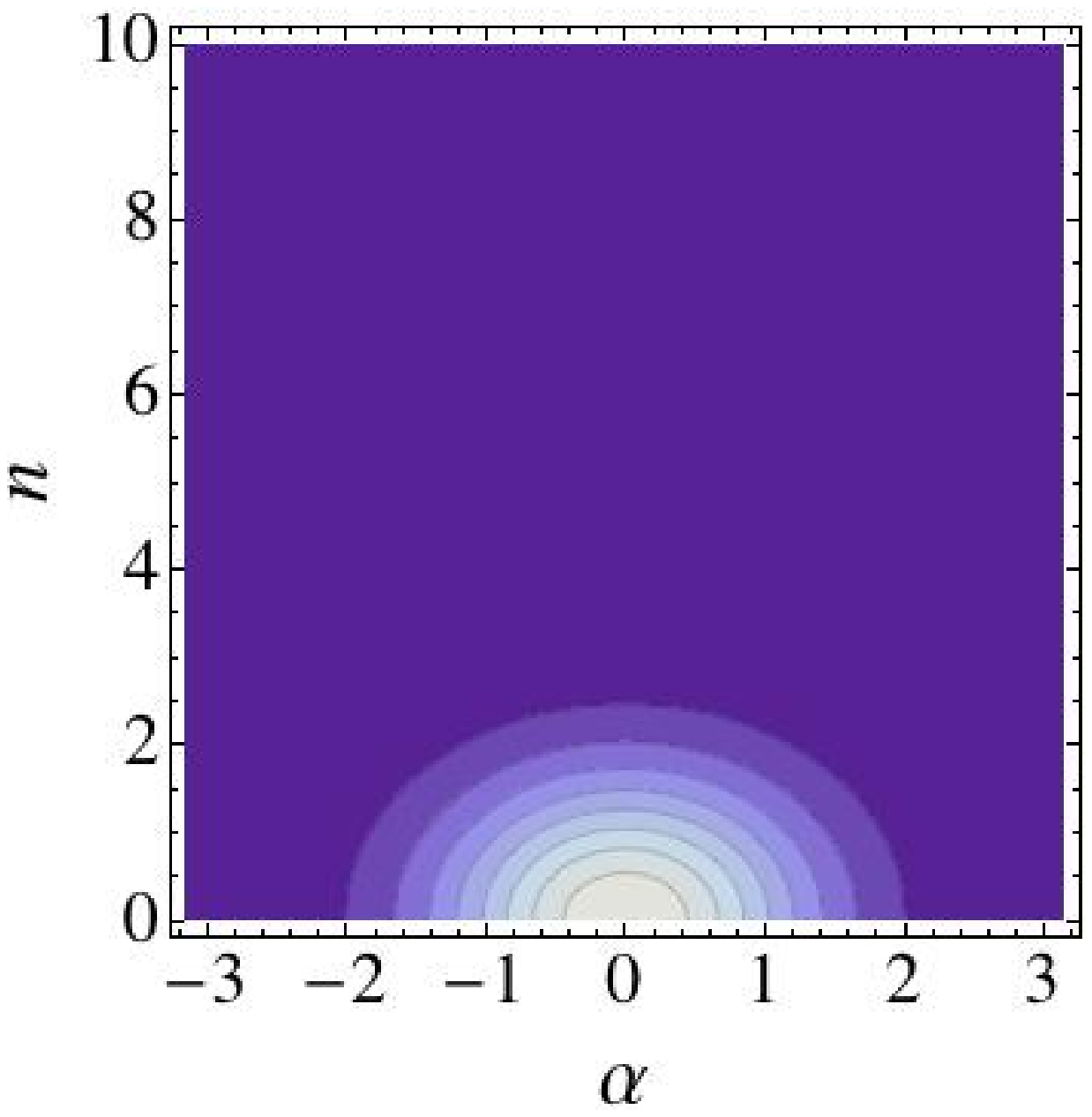}
\hspace{0.04\linewidth}
\includegraphics[width=0.46\linewidth]{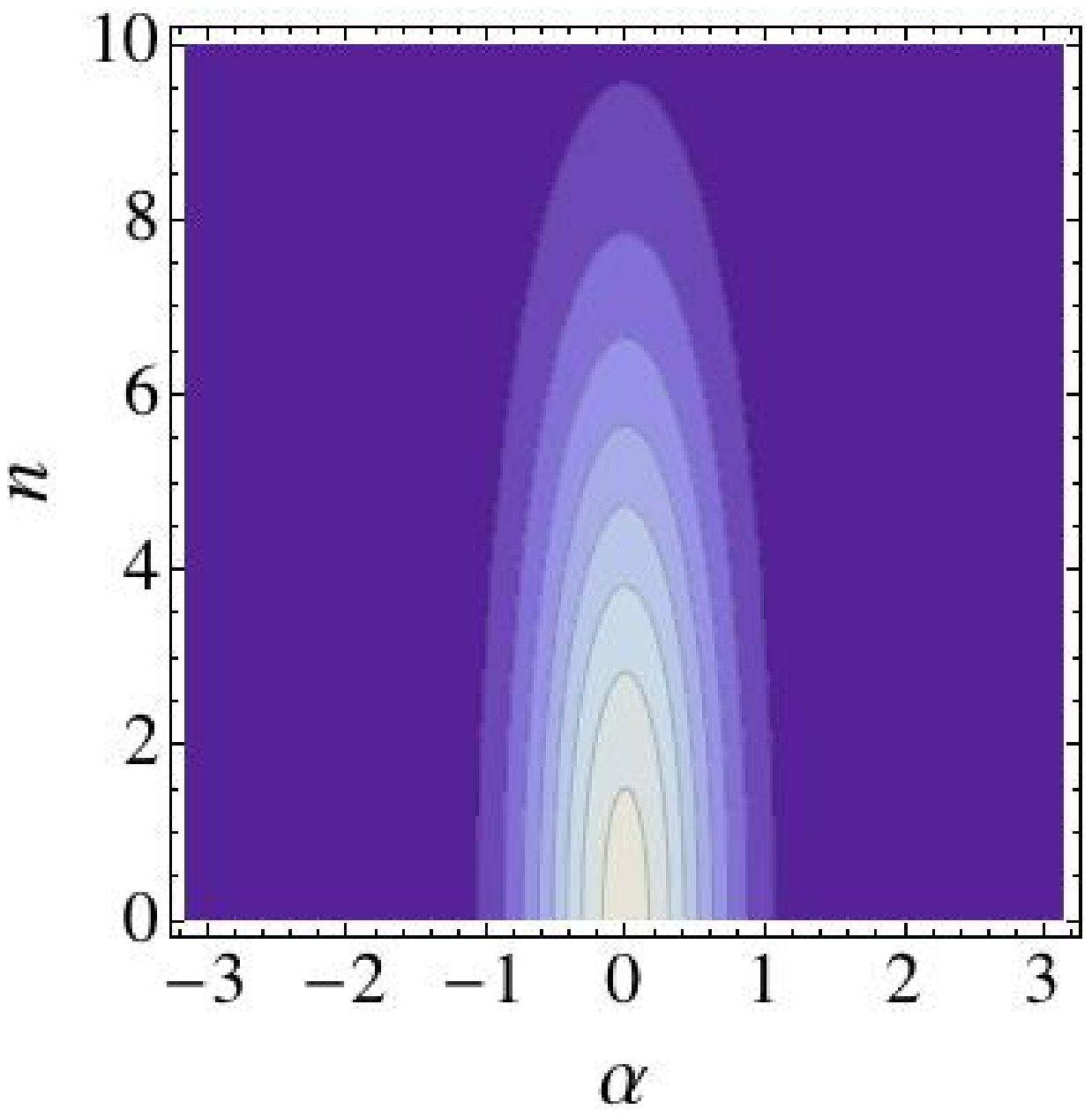}}
\caption{Husimi function \eqref{eq:Halphar} for the inflationary reheating model at $\mu\tau=0.5,1.5$ and for $\Delta=2$. The horizontal and vertical axes are chosen as in Fig.\ \ref{fig5}.} 
\label{fig6}
\end{figure}

Finally, we obtain the Wehrl entropy:
\ba
S_{{\rm H},\Delta}(\tau) &=& - \int \frac{d\alpha}{2\pi}\sum_{n} H_\Delta \ln H_\Delta
\approx
\ln\frac{1+\tilde\pi(\tau)^2\Delta}{\sqrt{\tilde\pi(\tau)^2\Delta}} + \frac{1}{2}
\nn \\
&\buildrel{\tau\to\infty}\over{\longrightarrow}& 2\mu\tau + {\rm const.}
\label{eq:SHinf}
\ea
Once again, the rate at which the Wehrl entropy increases is given by the classical Lyapunov exponent $2\mu$. Taking several independent modes into account one obtains as before the sum of all positive Lyapunov exponents, i.e. the Kolmogorov-Sina\"i entropy.

\section{Relation to Other Definitions of Entropy}
\label{sec:Other}

\subsection{Thermal equilibrium}

We next discuss the relation between the Husimi-Wehrl entropy discussed in the previous section and the von Neumann entropy in thermal equilibrium. The simplest system to consider is a harmonic oscillator,
\be
{\cal H}=\frac{p^2}{2}+\frac12\omega^2x^2\ .
\ee
It is also a quite generally applicable, because a quantum field at weak coupling or low excitation can be considered as an ensemble of infinitely many harmonic oscillators. The occupation probability of the eigenstate $\ket{n}$ is given as,
\ba
w_n &=& e^{-n\beta\hbar\omega}/\Zbeta\ ,\\
\Zbeta &=& \sum_{n=0}^\infty e^{-n\beta\hbar\omega}
= (1-e^{-\beta\hbar\omega})^{-1}\ .
\ea
where $\beta=1/T$ denotes the inverse temperature, and $\Zbeta$ is the partition function.The eigenstate $\ket{n}$ corresponds
to the $n$-th excited state in quantum mechanics and to the $n$-particle state in a Fourier mode $\omega = E_p$ in the case of a field theory. The von Neumann entropy is calculated to be 
\ba
S_{\rm vN} &\equiv& - \sum_{n=0}^\infty w_n \ln w_n 
	= \frac{\beta\hbar\omega}{e^{\beta\hbar\omega}-1}
	-\ln(1-e^{-\beta\hbar\omega})
\label{eq:S_vN}
\\
&=& -\bar{n}\ln\bar{n}+(\bar{n}+1)\ln(\bar{n}+1) \ .
\ea
In the second line, the average number of excited quanta (or the average number of particles in the field theory) is used,
\be
\bar{n}=\frac{1}{\Zbeta}\sum_{n=0}^\infty n w_n
= \frac{1}{e^{\beta\hbar\omega}-1}\ .
\ee

The thermal Wigner function is given by \cite{Hillery:1983ms}
\be
W(z)=B_\beta \exp(-B_\beta\,\bar{z}z) \ ,
\ee
where $B_\beta=2\tanh(\beta\hbar\omega/2)=1/(\bar{n}+1/2)$, and we have applied a complex representation of $(x,p)$ as
$z=(\omega x + ip)/\sqrt{2\hbar\omega}$. Although the Wigner function is, in general, not positive definite, it is positive around the thermal equilibrium and can thus be used to define an entropy, which we here call the Wigner-Wehrl entropy $S_{\rm W}$. By using the phase space measure,
$d\Gamma = dx dp /2\pi\hbar = dz d\bar{z}/\pi 
(\equiv d(\mathrm{Re}\,z) d(\mathrm{Im}\,z)/\pi)
= d(\bar{z}z) d\theta/2\pi$,
the Wigner-Wehrl entropy is obtained as,
\ba
S_{\rm W}&=&-\int d\Gamma W(z)\ln W(z) 
= - \int \frac{d(\bar{z}z)d\theta}{2\pi} B_\beta e^{-B_\beta\bar{z}z}
	\ln(B_\beta e^{-B_\beta\bar{z}z})
\nn\\
&=&1-\ln B_\beta = 1 + \ln \left(\bar{n}+\frac12\right) \ ,
\label{eq:S_WW}
\ea
where $\theta$ denotes the argument of $z$.

The Husimi function is the expectation value of the density matrix with the coherent state defined as,
\be
\ket{z}=e^{-\bar{z}z/2}\exp(z\hat{a}^\dagger)\ket{0}
=e^{-\bar{z}z/2}\sum_{n=0}^\infty \frac{z^n}{\sqrt{n!}}\,\ket{n}\ ,
\ee
where we have chosen the natural width $\Delta=\omega$ in Eq.~(\ref{eq:aDelta}) for simplicity. The Husimi function in thermal equilibrium is given as,
\ba
H(z)&=&\bra{z}\hat{\rho}_\mathrm{th}\ket{z}
  =\frac{e^{-\bar{z}z}}{\Zbeta}\,
  \sum_{n=0}^\infty \frac{(\bar{z}z)^n}{n!}\,e^{-n\beta\hbar\omega}
  \nn\\
  &=& \frac{1}{\Zbeta}\,\exp\left[
  	-\bar{z}z\left(1-e^{-\beta\hbar\omega}\right)
	\right]
   = A_\beta \exp\left(-A_\beta\bar{z}z\right) \ ,
\ea
where $\hat{\rho}_\mathrm{th}=\sum_{n=0}^\infty \ket{n}\,w_n\,\bra{n}$ denotes the density matrix in thermal equilibrium, and $A_\beta=1-e^{-\beta\hbar\omega}=1/(\bar{n}+1)$. The Husimi-Wehrl entropy $S_{\rm H}$ then is obtained as
\ba
S_{\rm H} &=& - \int d\Gamma H(z) \ln H(z)
= - \int \frac{d(\bar{z}z)d\theta}{2\pi} A_\beta e^{-A_\beta\bar{z}z}
	\ln(A_\beta e^{-A_\beta\bar{z}z})
\nn\\
&=& 1 - \ln A_\beta = 1 + \ln (\bar{n}+1)
\label{eq:S_HW} \ .
\ea

\begin{figure}[htb]
\centerline{
\includegraphics[width=0.68\linewidth]{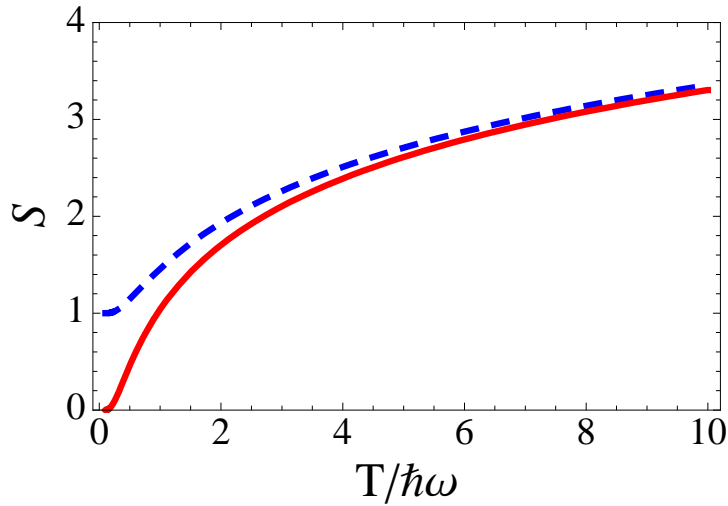}}
\caption{von Neumann entropy $S_{\rm vN}$ shown as solid line and Husimi-Wehrl entropy $S_{\rm H}$ shown as dashed line for the harmonic oscillator at thermal equilibrium as function of the temperature $T$, scaled by the characteristic oscillator energy $\hbar\omega$. }
\label{fig3}
\end{figure}

The Wigner-Wehrl and Husimi-Wehrl entropies, $S_{\rm W}$ and $S_{\rm H}$ in Eqs.~(\ref{eq:S_WW}) and (\ref{eq:S_HW}), respectively, differ from the von Neumann entropy, $S_{\rm vN}$ in Eq.~(\ref{eq:S_vN}), at low temperatures ($T \lesssim \hbar\omega$), while all three entropy definitions approach the same asymptotic value, $1+\ln\bar{n}\simeq 1-\ln\beta\hbar\omega$, at high temperatures ($T \gg \hbar\omega$). 
As discussed in Ref.~\citen{Bengtsson}, the Wehrl entropies are always greater
than or equal to the von Neumann entropy (Wehrl-Lieb theorem), 
and in the present case, we have shown that the difference is at most one unit
as far as we adopt the natural width.
The temperature dependence of the von Neumann entropy and the Husimi-Wehrl entropy for the harmonic oscillator are plotted in Fig.~\ref{fig3}. Since the average occupation number of a quantum field mode in thermal equilibrium grows with temperature as $\bar{n} \approx T/\hbar\omega \gg 1$ (for $T \gg \hbar\omega$) we can surmise that, in general, $S_{\rm H}$ will be a good approximation to the physical entropy for field modes that are highly occupied. In the application to the equilibration phase of relativistic heavy ion collisions, which we will discuss in the next section,  this condition is satisfied, because in the domain of interest, the so-called ``glasma'' phase, the average occupation number of field mode is $\bar{n} \approx 1 / \alpha_s \approx 3,$ where the difference between $S_H$ and $S_vN$ is less than 10 percent.

The difference between the Husimi-Wehrl entropy and the von Neumann entropy is caused by the energy fluctuation in the coherent state. In order to demonstrate this point, we consider the case where the partition function is given in the coherent state representation as follows~\cite{Ohnishi:1996ic},
\ba
\Zbeta &=& \int d\Gamma \bra{z}e^{-\beta{\cal H}}\ket{z}
 = \int d\Gamma \exp\left[
 	- \int_0^\beta\,d\beta' {\cal H}_{\beta'}(z)
 	\right]
\ ,\\
{\cal H}_\beta(z)
&=&-\frac{\partial}{\partial\beta}\ln\bra{z}e^{-\beta{\cal H}}\ket{z}
=\bra{z}{\cal H}\,e^{-\beta{\cal H}}\ket{z}/
		\bra{z}e^{-\beta{\cal H}}\ket{z}
\nn\\
&=& \bra{z}{\cal H}\ket{z} - \beta\sigma^2_{\cal H}(z) + {\cal O}(\beta^2)
\ ,
\ea
where ${\cal H}_\beta(z)$ is the thermally distorted expectation value of the Hamiltonian, and $\sigma^2_{\cal H}(z)$ represents the energy variance $\sigma^2_{\cal H}(z)=\bra{z}{\cal H}^2\ket{z}-\bra{z}{\cal H}\ket{z}^2$. From general thermodynamic relations, the (von Neumann) entropy is calculated as
\ba
S &=& - \frac{\partial F}{\partial T}
   =  \frac{\partial}{\partial T}\left(T \ln \Zbeta\right)
   = - \beta\frac{\partial}{\partial\beta}\ln\Zbeta + \ln \Zbeta
\nn\\
  &=& \int d\Gamma\,
	\frac{\bra{z}e^{-\beta{\cal H}}\ket{z}}{\Zbeta}
	\left[
  	\beta\,{\cal H}_\beta(z)
	+\ln\Zbeta
	\right] \ .
\label{eq:S_OR}
\ea
The difference from the Husimi-Wehrl entropy is evaluated to be,
\ba
S_{\rm H} - S 
  &=& \int d\Gamma\,H(z)\,\left[
  		\int_0^\beta d\beta'\, {\cal H}_{\beta'}(z)
  		- \beta{\cal H}_\beta(z)
  		\right]
\nn\\
&\simeq&
  \int d\Gamma\,H(z)\,\left[
  		\frac12\beta^2\sigma^2_{\cal H}(z)
		+ {\cal O}(\beta^3)
  		\right] \ .
\ea
This relation shows that Husimi-Wehrl entropy agrees with the von Neumann entropy when the energy fluctuation is much smaller than the temperature. Systematic corrections are possible when we can evaluate the expectation value of ${\cal H}^n$. Specifically, in the case of the harmonic oscillator, the evolution of ${\cal H}(z)$ with $\beta$ and the energy variance are known as, ${\cal H}_\beta(z)=\bra{z}{\cal H}\ket{z}\,\exp(-\beta\hbar\omega)$ and $\sigma^2_{\cal H}(z)=(\hbar\omega)^2\bar{z}z$, allowing us to obtain explicit correction terms which exactly reproduce the von Neumann entropy, if this is desired.

\subsection{Dynamical versus kinetic regime}

For systems which permit a quasi-particle description, the Kadanoff-Baym formalism \cite{Kadanoff_Baym} can be used to derive the kinetic theory from quantum field theory. The derivation usually starts from the single-particle Wigner function
\be
W_{\rm sp}(p,x;t) = \int du \; e^{\frac{i}{\hbar}pu} \langle \Phi(x - \frac{u}{2}) \Phi(x + \frac{u}{2}) \rangle ,
\label{eq:Wsp}
\ee
and performs the Moyal (gradient) expansion to define the phase space distribution $f(x,p,t)$. One can then define the kinetic entropy via the standard phase space integral (see e.~g.\ \cite{Kita:2006}):
\be
S_{\rm kin} = \int \frac{d^3x d^3p}{(2\pi \hbar)^3} \, [ - f \ln f + (1 + f)\ln (1+ f) ] .
\ee
(For fermions, the plus signs are replaced by minus signs in the second term in square brackets.) The kinetic entropy so defined agrees with the von Neumann entropy in equilibrium and satisfies the $H$-theorem for processes relaxing to or close to the thermal equilibrium state. An extension to include memory effects in the kinetic equations is also possible \cite{Ivanov:1999tj}. The Kadanoff-Baym formalism was recently applied by Nishiyama to study entropy production in the kinetic regime of the scalar $\Phi^4$-theory \cite{Nishiyama:2008zw}. Its extension to nonabelian gauge fields can be found in the review of Elze and Heinz \cite{Elze:1989un}; the application to relativistic nuclear collisions was explored by Gelis {\em et al.} \cite{Gelis:2007pw}.

The Kadanoff-Baym formalism makes various approximations, some of which break time-reversal invariance, in order to derive a kinetic equation: ($i$) A truncation of the Feynman diagrams at the two-particle irreducible (2PI) level is imposed. This truncation implies an information loss, because two-and more-particle correlations in the quantum state are neglected. ($ii$) Since it is impossible to include all the diagrams of the 2PI type, one must further make a skelton expansion and take only diagrams with a limited number of loops. ($iii$) Finally, the gradient expansion is made to break the space-time nonlocality of the quantum theory. This amounts to a Markovian approximation.

Here, on the other hand, we are interested in the passage from the dynamical (Liouville) regime of the quantum field theory to the kinetic regime. This problem arises when the system of interest starts in a highly excited, but still nearly coherent quantum state and then evolves toward equilibrium. In such situations, one cannot neglect the many-particle correlations which are encoded in coherent field configurations, because it is precisely the loss of coherence of such configurations due to their own nonlinear dynamics, which leads to the initial growth of entropy. The Kadanoff-Baym approach to kinetic theory is not applicable to this domain. 

It is not clear at all, whether a unique definition of the concept of entropy can be given, which permits the description of the transition from pure quantum state to thermal equilibrium and applies to all possible systems. The definition of entropy is firmly linked to the concept of coarse graining of the quantum state and thus depends on the specific chosen procedure. The Husimi distribution is one possible choice; it is motivated by the idea that coherent field configurations play an important role during the transition from the (nearly) pure quantum state to the kinetic regime. We emphasize, however, that we do not assign special importance to the absolute value of the Husimi-Wehrl entropy $S_{\rm H}$ --- since its precise value depends on the arbitrary smearing parameter $\Delta$, this would be difficult to justify --- but only to its growth rate $dS_{\rm H}/dt$, which is independent of $\Delta$ in the long-time limit. It would be interesting to study how the growth rate of the Husimi-Wehrl entropy compares with the growth rate of the kinetic entropy, as a system enters the kinetic regime. We leave this question to a future study.

\section{Relativistic heavy ion collisions}
\label{sec:RHIC}

Much experimental evidence exists showing that a thermal quark-gluon plasma is formed in collisions between two heavy nuclei at center-of-mass energies on the order of 100 GeV per nucleon \cite{RHIC}. The total entropy per unit of rapidity, $dS/dy$, of the final state in such a reaction of two Au nuclei has been estimated to be approximately 5,000 \cite{Pal:2003rz,Muller:2005en}. How, and how rapidly, is this entropy produced? Experimental observations, in combination with hydrodynamics simulations, suggest that the thermalization time is very short, of order 1 fm/c or less \cite{Heinz:2001xi}. A fraction of the final entropy can be produced simply by decoherence of the initial quantum states of the colliding nuclei, on a time scale $\tau_\mathrm{dec} \sim 1/Q_s$, where $Q_s$ is the saturation scale of the nuclear parton distributions \cite{Muller:2005yu,deco1}. 

The remainder of the entropy can be generated at various stages of the heavy-ion collisions: One source which is certainly relevant are nonlinear interactions among the gluon fields liberated in the decoherence process, but the precise mechanism responsible for the entropy growth and thermalization is still unclear. Suggested processes include the Boltzmann cascades of partons \cite{Geiger:1991nj,Shin:2003yk,Xu:2004mz}, the nonlinear dynamics of classical color fields \cite{Krasnitz:2002mn,Krasnitz:2003jw,Lappi:2003bi}, the decay of unstable chromo-magnetic fields \cite{Fujii:2008dd,Iwazaki:2008xi}, and plasma instabilities in the longitudinally expanding matter \cite{Mrowczynski:2005ki,ALMY,RRS}. 
Another relevant source might be the viscosity terms in viscous hydrodynamics \cite{Kharzeev:2007wb,Karsch:2007jc,Dumitru:2007qr,Torrieri:2008ip,Fries:2008ts}. The hadronic phase and processes like jet fragmentation \cite{Isse:2007pa} are promising sources. Here we concentrate, however, on the initial phase of the heavy-ion collision.

All  approaches cited above are based on classical or quasi-classical pictures, in which the production of entropy is mostly assumed (as in parton cascades) or based on plausibility arguments (as in simulations of classical Yang-Mills fields), but not rigorously calculated. As far as we know, no comprehensive formalism has ever been proposed which 
would, even in principle, describe the calculation of entropy production starting from the initial quantum states of the colliding nuclei. Fukushima {\em et al.} \cite{Fukushima:2006ax} analyzed the quantum fluctuations around classical glue fields in the earliest phase of the collision, but did not address the problem of entropy production. Their analysis 
is also problematic in our context, because it assumes that the collision time is exactly zero.

In principle, the Husimi function approach described in the previous sections provides the missing link in the formulation of a comprehensive approach to entropy production in nuclear collisions. The lattice regularized Yang-Mills 
equations describing the dynamics of classical color fields were found to be strongly chaotic \cite{Matinyan:1986nw,Muller:1992iw,Biro:1993qc}, and the KS-entropy of the classical Yang-Mills field was shown to be a thermodynamically extensive quantity 
\cite{Bolte:1999th}. The general formulation of the dynamics of quantum fields based on the Wigner functional has been developed by Mrowczynski and one of the present authors \cite{Mrowczynski:1994nf}. An extension of the classical lattice Yang-Mills theory to include Gaussian quantum fluctuations around the classical link variables was proposed by 
Gong {\em et al.} \cite{Gong:1993fz}. Based on these developments one can construct a Husimi functional for the lattice Yang-Mills field, using either the Wigner functional or the Gaussian variational method, and thereby link the classical 
calculation of the KS-entropy density to the entropy density growth rate of the quantized gluon field.

\section{Summary}
\label{sec:Summary}

We have proposed a method to determine the rate of entropy production for quantum field theoretical systems like colliding nuclei or the early universe. This method employs the Husimi function to implement the amount of coarse graining required by the uncertainty principle. For the  entropy we adopt the definition of Wehrl. We then find that for large enough times the rate of entropy growth is given by the analog of the classical Kolmogorov-Sinai entropy. This observation agrees with results from various discussions found in the literature. We illustrated these general ideas for a few concrete examples.  In Sections \ref{sec:GrowthRate} and \ref{sec:RollOver} we analyzed the situation for a scalar field in an inverted harmonic oscillator potential. The standard toy-model for pre-heating after cosmological inflation was analyzed in Section \ref{sec:BigBang}. 
In Section \ref{sec:Other} we compared the Wigner-Wehrl and Husimi-Wehrl entropy with other definitions of entropy, such as the von Neumann entropy and kinetic
entropy.

Based on these observations and examples we have shown that it is possible to define the conceptual framework for a calculation of the entropy growth rate in heavy ion collisions which rests on a much firmer theoretical basis than previously. In Section \ref{sec:RHIC} we outlined the steps necessary to carry out such a calculation. Its realization will require a prolonged and extensive effort. The initial conditions for  the gluon Wigner distribution must be derived from the gluon wave functions of the colliding nuclei, which is only known approximately within certain models (see {\em e.~g.}) \cite{Kovchegov:1996ty}). The time evolution of the Wigner and Husimi functions for the Yang-Mills field will be strikingly different from that of the simple cases analyzed here, in which different modes did not couple. The time evolution as well as the Husimi transform will have to be done numerically, due to the complexity of the problem. The accuracy and precision of the applied numerical methods will have to be tested carefully. We should confirm, {\em e.g.}, that the  growth rate of the entropy density becomes independent of the number of lattice points used to discretize space for the quantized Yang-Mills field. However, all of these problems are technical rather than conceptual and will hopefully be resolvable in due course.

\section*{Acknowledgments}

This work was initiated at the Workshop on {\em Entropy Production before QGP} at the Yukawa Institute of Theoretical Physics during August 2008. It was supported in part by the YITP and grants from the U.S. Department of Energy (DE-FG02-05ER41367) and the Bundesministerium f\"ur Bildung und Forschung. T.K. and A.O. are partially supported by a Grant-in-Aid for Scientific 
Research by the Ministry of Education, Culture, Sports, Science and Technology (MEXT) of Japan (nos. 20540265, 17070002 and 19540252) and by the Grant-in-Aid for the global COE program 'The Next Generation of Physics, Spun from Universality and Emergence' from MEXT. A.S. acknowledges support as visiting professor at YITP. A.S. and B.M. thank the members of the YITP for their hospitality and we all thank Toru Takahashi for most helpful discussions. The authors also thank Arjendu Pattanayak for useful discussions and Tamas Bir\'o for helpful comments on the mansucript.

%


\end{document}